%% file: main.tex
\documentclass[sigconf]{acmart}

\usepackage{booktabs} 

\usepackage{graphicx}
\usepackage{caption}
\usepackage{caption}
\usepackage{natbib}
\usepackage{amsthm,amsmath}
\usepackage{mathrsfs}
\usepackage{soul}
\usepackage{url}
\usepackage{enumerate}
\usepackage[utf8]{inputenc}
\usepackage{mathtools}
\usepackage{algorithm}
\usepackage{algorithmic}
\usepackage{subfigure}   
\usepackage{cancel}
\usepackage{comment}
\usepackage{ctable}
\usepackage{mathtools}
\usepackage{balance}
\usepackage{threeparttable}  
\usepackage{booktabs}
\usepackage{multirow}
\usepackage{bm}
\usepackage{enumitem}
\usepackage{amsthm}
\usepackage[switch]{lineno} 
\usepackage{xcolor,colortbl}
\theoremstyle{definition}   
\newtheorem{Definition}{Definition} 

\newtheorem{Problem}{Problem}
\newtheorem{Example}{Example}

\newcommand{\xhdr}[1]{\vspace{2mm}{\noindent\bfseries #1}.}
\usepackage{balance}
\usepackage{geometry}
\usepackage{pifont}
\newcommand{\cmark}{\text{\ding{51}}}

\acmPrice{}

\copyrightyear{2022}

\acmYear{2022}
\renewcommand\footnotetextcopyrightpermission[1]{}
\setcopyright{none}

\begin{document}
\title[]{HIST: A Graph-based Framework for Stock Trend Forecasting via Mining Concept-Oriented Shared Information}

\author{Wentao Xu\textsuperscript{1}*, Weiqing Liu\textsuperscript{2}, Lewen Wang\textsuperscript{2}, Yingce Xia\textsuperscript{2}, Jiang Bian\textsuperscript{2}, Jian Yin\textsuperscript{1}, Tie-Yan Liu\textsuperscript{2}}
\thanks{*Work done while Wentao Xu was an intern at Microsoft Research}
\affiliation{\textsuperscript{1}Sun Yat-sen University}
\affiliation{\textsuperscript{2}Microsoft Research}
\email{{xuwt6@mail2,issjyin@mail}.sysu.edu.cn}
\email{{weiqing.liu, lewen.wang, yingce.xia, jiang.bian, tyliu}@microsoft.com}
\renewcommand{\authors}{Wentao Xu, Weiqing Liu, Lewen Wang, Yingce Xia, Jiang Bian, Jian Yin, Tie-Yan Liu}
\renewcommand{\shortauthors}{Xu et al.}

\input{abstract}

%
%



\maketitle

\input{introduction}
\input{related}

\input{preliminaries}

\input{method}
\input{experiment}
\input{conclusion}
\bibliographystyle{ACM-Reference-Format}
\bibliography{HIST}
\input{appendix}

\end{document}

%% file: abstract.tex
\begin{abstract}
Stock trend forecasting, which forecasts stock prices' future trends, plays an essential role in investment. The stocks in a market can share information so that their stock prices are highly correlated.  Several methods were recently proposed to mine the shared information through stock concepts (e.g., technology, Internet Retail) extracted from the Web to improve the forecasting results. 
However, previous work assumes the connections between stocks and concepts are stationary, and neglects the dynamic relevance between stocks and concepts, limiting the forecasting results.
Moreover, existing methods overlook the invaluable shared information carried by hidden concepts, which measure stocks' commonness beyond the manually defined stock concepts.
To overcome the shortcomings of previous work, we proposed a novel stock trend forecasting framework that can adequately mine the concept-oriented shared information from predefined concepts and hidden concepts. The proposed framework simultaneously utilize the stock's shared information and individual information to improve the stock trend forecasting performance. 
Experimental results on the real-world tasks demonstrate the efficiency of our framework on stock trend forecasting. The investment simulation shows that our framework can achieve a higher investment return than the baselines. 
\end{abstract}

%% file: introduction.tex
\section{Introduction}
\label{sec:intro}
The stock market is one of the most profitable investment channels in the real world. To pursue high yield by investing in it, stock trend forecasting has attracted increasing attention in recent years as a fundamental component of many complex investment strategies.
Many of existing efforts~\cite{akita2016deep,patel2015predicting,ticknor2013bayesian} assume that the prices of different stocks are independent with each other and build the forecasting model merely based on information related to each stock, such as time series of historical stock price and volume (e.g., \textit{opening price}, \textit{closing price}, \textit{highest price} and \textit{trading volume}). 

\begin{figure}[t]
	\begin{center}
		\includegraphics[width=0.9\columnwidth]{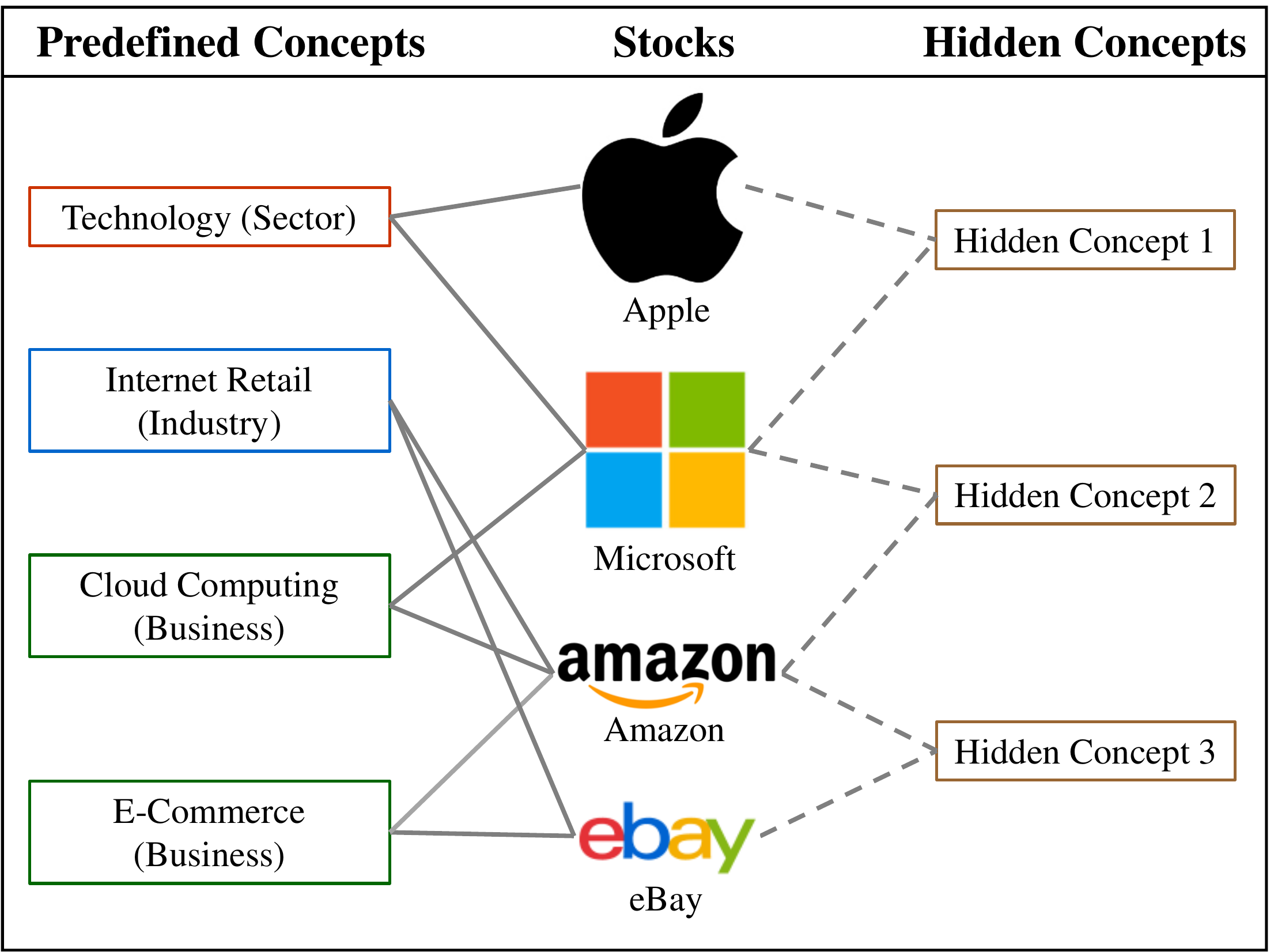}
	\end{center}
		\vspace{-0.1in}
	\caption{Some examples of stocks, predefined stock concepts and hidden stock concepts.}
	\vspace{-0.1in}
	\label{fig:concept_example}
\end{figure}

However, in practice, the price trends of different stocks tend to be highly correlated with each other when these stocks bear the shared concept. Such concepts are usually extracted from public company information on the Web, and can be specified by various dimensions, such as sector, industry, business, etc. For instance, most of the stocks under the high-tech concept have been sharing a similar bull trend along with the rapid development of information technology. For another example, many listed companies related to the concept of medicine have experienced drastic stock price surges after the outbreak of COVID-19 pandemic. 
Figure~\ref{fig:concept_example} illustrates some examples of stocks with their corresponding predefined concepts. 
Many recent studies have turned their eyes to utilizing such information in stock trend forecasting by recognizing the valuable information in stock concepts.
For example, some straightforward methods~\cite{rapach2019industry} directly used the predefined stock concepts as input features of the linear forecasting model. 
In addition, some others~\cite{feng2019temporal,kim2019hats,matsunaga2019exploring,li2020modeling} used the same predefined concepts to form up relations between two stocks and leveraged the Graph Neural Network (GNN), whose edges are defined by conceptual relations, to build a more accurate stock trend forecasting model.

While these recent studies have revealed the potential of stock concepts in boosting stock trend forecasting, they are still enduring some limitations that restrain them from fully leveraging the value of stock concepts.
In the first place, when building the GNN model with the predefined concepts as connections between stocks, most previous studies assume that those connections are stationary such that information propagation through them follows the same pattern. However, in the real financial market, one stock could yield a dynamic relevance degree to various concepts. For example, as shown in Figure~\ref{fig:concept_example}, Amazon has two predefined concepts and shares the same business concept `cloud computing' with Microsoft and the other `e-commerce' with eBay. Apparently, during the lockdown period caused by the COVID-19 pandemic, the rising stock price trend of Amazon is mainly due to its conceptual bond to surging `e-commerce' growth\footnote{https://www.forbes.com/sites/sergeiklebnikov/2020/07/23/5-big-numbers-that-show-amazons-explosive-growth-during-the-coronavirus-pandemic} rather than `cloud computing'. 
However, most of the existing GNN-based methods overlook differentiating the information propagation through various concepts. 

Moreover, since most existing studies merely leverage the concepts predefined by human experts, they will miss some hidden concepts. Even some emerging important concepts may not be able to be promptly included in the modeling. In the meantime, some recently-listed companies may not benefit from the predefined concepts until the corresponding conceptual connections are complemented.
For instance, a Personal Protective Equipment company may accidentally have shared information and similar future trend with an e-commerce company because of a sudden outbreak of a pandemic (a hidden concept representing the pandemic-related companies) but will lose this correlation after the pandemic period.

To address these limitations, we proposed a novel graph-based framework to mine the concept-oriented s\underline{h}ared \underline{i}nformation for \underline{s}tock \underline{t}rend forecasting (HIST). 
The design of our HIST mainly follows the below two principles:
\begin{enumerate}[leftmargin=1.5em]
	\item[{(1)}] To differentiate a stock's information propagating to related stocks with different predefined concepts, we explicitly learn the dynamic representation of various concepts by jointly modeling the relevance degrees and aggregating information from corresponding stocks. Specifically, we construct a stock-concept bipartite graph to properly extract the dynamic representations of predefined concepts and propagate them to stocks with similar representation, regardless of whether the predefined concept covers the stock.
	
	\item[{(2)}] To discover the hidden concepts and the corresponding concept representations, we first introduce a doubly residual architecture~\cite{oreshkin2019n} upon the stock-concept bipartite graph to extract the remaining information of each stock after the shared information of predefined concepts is filtered out. Subsequently, based on the stocks' remaining information, a simple yet effective graph algorithm is designed to dynamically detect hidden concepts of each time-step and construct their hidden concepts' representations. 
\end{enumerate}
After the information propagation via mined dynamic hidden concepts, we extract each stock's individual information by further filtering out the shared information of dynamic hidden concepts.
Finally, we utilize the shared information of predefined concepts, the shared information of hidden concepts, and each stock's individual information to simultaneously forecast the stock price trend.

We evaluate our framework on real-world stock data, and the experimental results show that our framework can outperform a couple of baselines in terms of various evaluation metrics. Moreover, we simulate the stock investment using a simple but widely-used trading strategy, and the results show that our framework can achieve a higher investment return than all baselines.
We also conduct additional analysis to investigate the effects of different components in our framework and visualize the mined hidden concepts to further reveal the advantages of HIST.

The main contributions of this paper include: 
\begin{itemize}[leftmargin=1.5em]
	\item  We proposed a new framework to mine the stocks' shared information, including the shared information of predefined concepts and hidden concepts. The shared information we mined can reflect the valuable indication of the stock's future trends with shared commonness.
	\item Our proposed framework can improve the stock trend forecasting performance by utilizing the dynamically shared information on the predefined and hidden concepts, and the individual information of each stock, simultaneously.
	\item We can discover some significant and valuable hidden concepts of stocks through our framework.
	\item We conducted the experimental evaluation and investment simulation on the real-world data, and the results verified our HIST framework's validity.
\end{itemize}

%% file: related.tex
\section{Related Work}
\label{sec:related}
Stock trend forecasting has attracted soaring attention because it is vital in stock investment. 
This section will introduce two representative categories of stock trend forecasting methods: technical analysis and event-driven stock trend forecasting methods. 
\subsection{Technical Analysis}
The technical analysis~\cite{edwards2018technical} predicts the stock trend based on the historical time-series of market data, such as trading price and volume. 
It aims to discover the trading patterns that we can leverage for future predictions.
We can further divide the technical analysis methods into the \textbf{single-stock methods} and the \textbf{cross-stock methods}. 
The single-stock methods only use each stock's information to forecast the stock trend, and the cross-stock methods consider the cross-stock relationships between stocks when they forecast the stock price trend.

\xhdr{Single-stock Methods}
For the single-stock methods, Autoregressive (AR)~\cite{li2016stock}, and ARIMA~\cite{ariyo2014stock} models are the most widely used model in this direction, which are both for linear and stationary time-series. However, the non-linear and non-stationary nature of stock prices limits the applicability of AR and ARIMA models. 
Some studies~\cite{ticknor2013bayesian,patel2015predicting} attempted to apply deep neural networks to catch the market trend's intricate patterns with the recent rapid development of deep learning. 
To further model the long-term dependency in time series, recurrent neural networks (RNN), especially Long Short-Term Memory (LSTM) network~\cite{hochreiter1997long}, had also been employed in financial predition~\cite{rather2015recurrent,gao2016stock,akita2016deep,bao2017deep}. Specifically,~\cite{zhang2017stock} proposed a new State Frequency Memory (SFM) recurrent network to discover the multi-frequency trading patterns for stock price movement prediction;~\cite{li2019multi} presented a multi-task recurrent neural network with high-order Markov random fields (MRFs) to predict stock price movement direction;~\cite{feng2019enhancing} leveraged adversarial training to simulate the stochasticity during model training. However, each stock is not isolated; the single-stock methods can not utilize the cross-stock interactions between stocks to capture more information for stock trend forecasting.

\xhdr{Cross-stock Methods}
To mine the cross-stock shared information and improve the stock trend forecasting performance, many cross-stock methods~\cite{chen2018incorporating,feng2019temporal,kim2019hats,matsunaga2019exploring} leveraged the Graph Neural Networks~\cite{kipf2017semi,velickovic2018graph} to capture the relationships between different stocks. The~\cite{chen2018incorporating} and~\cite{feng2019temporal} utilize the graph convolutional networks (GCN) to capture the stocks' shareholder relations and industry relations, respectively. The~\cite{kim2019hats} propose a hierarchical attention network for stock prediction (HATS), which uses relational data for stock market prediction.
However, these existing cross-stock methods can not correctly utilize the shared information of predefined concepts because they use stationary relations to aggregate information and neglect dynamic relevance between the stocks and predefined concepts.
Additionally, they also overlook the invaluable shared information of hidden concepts, limiting their stock trend forecasting performance. 

\subsection{Event-driven Stock Trend Forecasting}
The event-driven stock trend forecasting is another category of stock trend forecasting methods, aiming to mine the event information from various sources to forecast the stock price trend.
The sources of event information include the news~\cite{nassirtoussi2015text,ding2016knowledge,hu2018listening,deng2019knowledge,vargas2017deep,xu2021rest}, social media~\cite{si2013exploiting,yang2015twitter,zhou2016can,xu2018stock,DBLP:conf/cikm/WuZSW18}, and discussion board~\cite{li2014effect,nguyen2015sentiment,zimbra2015stakeholder}.
These methods can discover the implicit rules governing the stock price trend from the event information. 
Nevertheless, the event-driven methods highly rely on the event data. The sparse and irregular event date would reduce the flexibility and performance of the event-driven stock trend forecasting models.

%% file: preliminaries.tex
\section{Preliminaries}
\label{sec:preliminaries}
This section will introduce some definitions in our work and the problem of stock trend forecasting.
\begin{Definition}
	\textbf{Stock concept}.
	The predefined stock concepts are some human-defined concepts to a stock, such as the stock's sector, industry and main businesses. The hidden stock concepts are some hidden concepts that human experts do not pre-define and reflect some similar stock price trend among stocks under the same hidden concept.
\end{Definition}

\begin{Example}
	In Figure~\ref{fig:concept_example}, there are four stocks: Apple, Microsoft, Amazon, and eBay; four predefined concepts of stocks: Technology (Sector), Internet Retail (Industry), Cloud Computing  (Business), and E-Commerce (Business). 
	There are also three hidden stock concepts among stocks.
	The stock Microsoft has the predefined stock concepts Technology (Sector) and Cloud Computing (Business), and the hidden stock concept Hidden concept $1$ and Hidden concept $2$.
\end{Example}

\begin{Definition}
	\textbf{Stock Price Trend.} 
	Many previous work~\cite{hu2018listening,xu2021rest} define the stock price trend as the future change rate of the stock price. Following these settings, we define the stock price trend of stock $i$ at date $t$ as the stock price change rate of the next day:
	\begin{linenomath}
		\begin{align}
			d_i^t= \dfrac{Price_i^{t+1} - Price_i^t}{Price_i^t},
			\label{eq:stock-trend}
	\end{align}\end{linenomath}
\end{Definition}

\noindent where $Price_i^t$ could be specified by different values, such as the \textit{opening price}, \textit{closing price} and \textit{volume weighted average price} (VWAP)~\cite{berkowitz1988total}, and we use the \textit{closing price} in our work.

\begin{Problem}
	\textbf{Stock Trend Forecasting.}
	Given the specific stock features (e.g., the historical stock price and volume, the textual information from news and social media) of stock $i$ at date $t$, the stock trend forecasting aims to forecast the stock price trend $d_i^t$.
\end{Problem}

%% file: method.tex
\section{Our HIST Framework}
\label{sec:method}

\subsection{Overview of Workflow}
\label{sec:overview}

We first introduce the general workflow of HIST. The inputs of date $t$ consists of the stock features $S^t = \{s_1^t, s_2^t, ...,s_n^t \}$ of $n$ stocks and $m$ predefined concepts $T_j$, $j\in\{1,2,\cdots,m\}$, and our goal is to forecast the future trend of each stock. Note each stock feature $s_i^t$ is a sequence of the historical raw feature (like opening price, closing prices, volume, etc). 
Figure~\ref{fig:framework} demonstrates the architecture of the HIST framework. 
There are three steps to use our framework:

\noindent({\bf Step-1}) For each stock feature $s_i^t$, we use a stock feature encoder to extract the temporal features of each stock. In our work, we use a $2$-layer GRU network for this purpose.

\noindent({\bf Step-2}) Three modules will sequentially process the features obtained in Step-1 and the predefined concepts:
\begin{enumerate}
	\item  The predefined concept module is built upon a graph neural network that extracts the shared information of stocks based on the predefined concepts. Note that the predefined concepts will be used in this step only (details in Section \ref{subsec:exctrapredefined}, Section \ref{subsec:Aggregating} and \ref{subsec:output}).
	\item The hidden concept module, which focuses on mining the hidden shared information beyond that carried by predefined concepts (details in Section \ref{subsec:extrahidden}, Section~\ref{subsec:Aggregating} and Section~\ref{subsec:output}).
	\item Individual information module, which processes the individual information that cannot be captured by the two kinds of shared information above (details in Section \ref{subsec:individual}).
\end{enumerate}
\noindent({\bf Step-3}) Finally, we feed the three types of information extracted in step 2 into a feed-forward network for prediction.

We introduce more details of the above steps:

\begin{figure}[t]
	\begin{center}
		\includegraphics[width=1\columnwidth]{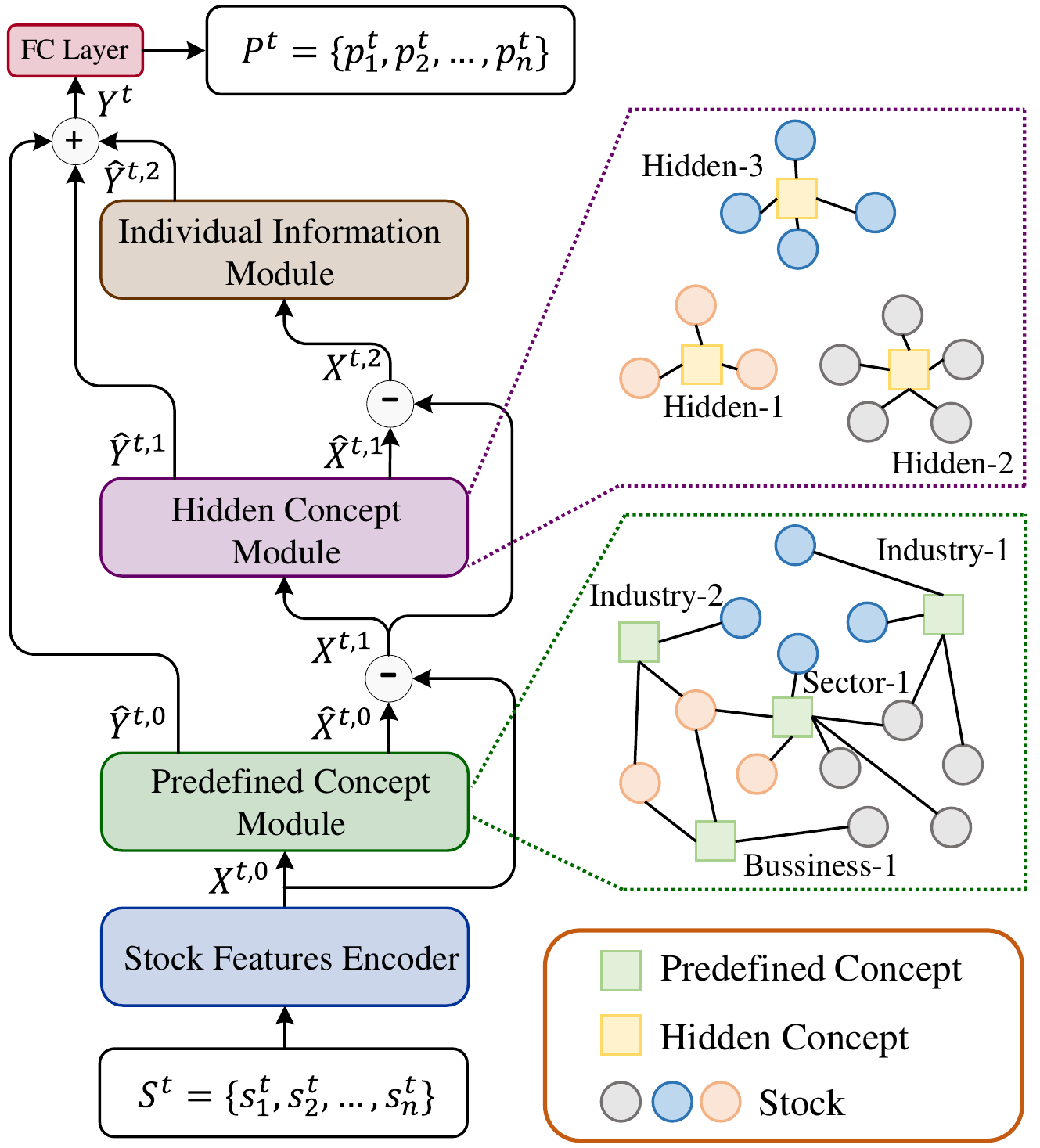}
	\end{center}
			\vspace{-0.1in}
	\caption{The overall architecture of the proposed HIST.}
	\vspace{-0.15in}
	\label{fig:framework}
\end{figure}

\xhdr{Stock Feature Encoder}
Let $s_i^t$ denote the feature of stock $i$ at date $t$, which is an $l$-dimensional historical stock prices and volume data.
Given the stock features $S^t = \{s_1^t, s_2^t, ...,s_n^t \}$ of $n$ stocks at date $t$, the stock feature encoder aims to encode the stock features to represent information for each stock in a low dimension space. Since Gated Recurrent Unit (GRU)~\cite{chung2014empirical} owns outstanding performance in capturing the long-term dependency, in this paper, we use a $2$-layer GRU network with hidden dimension $d$ as the stock feature encoder. Then we take the last hidden state of GRU's output as the initial embedding of the stocks at date $t$:
\begin{equation}
	x_i^{t,0} = \mathrm{GRU}\left(s_i^t\right),
\end{equation}
where $x_i^{t, 0}$ is the initial stock embedding of stock $i$. 
Then the stocks' initial embedding of all stocks at date $t$ is the matrix $X^{t,0}$, where the $i$-th row of $X^{t,0}$ is $x_i^{t, 0}$.

\xhdr{Doubly Residual Architecture}
As shown in Figure~\ref{fig:framework}, for Step-2 and Step-3, we follow \cite{oreshkin2019n} and use the doubly residual architecture. 
For convenience, we denote the predefined concept module, hidden concept module and individual module as module $0$, $1$ and $2$, respectively. For each module in Step-2, given a stock $i$, the input of the $j$-th module is $x_i^{t,j}$. Each module has two outputs, one forecast output $\hat{y}_i^{t,j}$ used for the final prediction, and one backcast output $\hat{x}_i^{t,j}$ used to remove the effect of current module for the next module. Note that $x_i^{t,j}=x_i^{t,j-1}-\hat{x}_i^{t,j-1}$, $j\ge1$, and $\hat{y}^t_i=\hat{y}^{t,0}_i+\hat{y}^{t,1}_i+\hat{y}^{t,2}_i$. 
For all stocks at date $t$, the matrix $X^{t,j}=\{x_{i}^{t,j}\}_{i=1}^n$ is the input of the $j$-th module, the matrices $\hat{X}^{t,j}=\{\hat{x}_{i}^{t,j}\}_{i=1}^n$ and $\hat{Y}^{t,j}=\{\hat{y}_{i}^{t}\}_{i=1}^n$ are the backcast ouput and forecast output of the $j$-th module, respectively. 

In HIST, the predefined concept module, the hidden concept module, and the individual information module are connected in the doubly residual structure. For the backcast residual branch, the predefined concept module's backcast output $\widehat{X}^{t,0}$ and hidden concept module's backcast output $\widehat{X}^{t,1}$ are designed for removing the shared information of predefined and hidden concepts from the next module's input, making the forecasting task of downstream module easier and facilitating more fluid gradient back-propagation. For the forecast residual, the forecast outputs $\widehat{Y}^{t,0}$, $\widehat{Y}^{t,1}$ and $\widehat{Y}^{t,2}$ are finally sum up to get the stock trends forecasting.

\xhdr{Stock Trend Prediction}
We feed the element-wise sum of the three modules' forecast output $\hat{y}^{t, 0}_i $, $\hat{y}^{t, 1}_i$, and $\hat{y}^{t, 2}_i$ into a fully-connected layer, and output the prediction $p_i^t$ for the stock future trend $d_i^t$ of stock $i$ at date $t$:
\begin{equation}
	p_i^t = W_p y_i^t  + b_p = W_p \left(\hat{y}_i^{t,0} + \hat{y}_i^{t,1}  + \hat{y}_i^{t,2} \right) + b_p.
\end{equation}

\xhdr{Following Organization} Section~\ref{subsec:exctrapredefined} to Section~\ref{subsec:output} will introduce details about the predefined concept module and hidden concept module. Both of these modules have three components: 1) extracting the concepts' representations from corresponding stocks; 2) aggregating the concepts' information to stocks according to the related concepts; 3) output of predefined/hidden concept module, including backcast output and forecast output.
We will introduce the first component of the predefined concept module and hidden concept module in Section~\ref{subsec:exctrapredefined} and Section~\ref{subsec:extrahidden}, respectively.
The predefined concept module and the hidden concept module share the same design for the second and third components. 
We will present the second component in Section~\ref{subsec:Aggregating} and the third component in Section~\ref{subsec:output}.
Besides, the Section~\ref{subsec:individual} and Section~\ref{sec:forecasting} describe the individual information module and training objective in details.

\subsection{Extracting the Predefined Concepts' Shared Information}
\label{subsec:exctrapredefined}
As mentioned in Section~\ref{sec:intro}, the degree of relevance between a stock and a concept may change dynamically. To model such dynamic connections, we propose to learn the concept representations to express the rich and temporal information of predefined concepts. In this paper, we propose to obtain concept representations from the information of corresponding stocks. To be specific, we first construct a stock-concept bipartite graph with the stocks and the predefined concepts in the predefined concept module. Then we extract the predefined concepts' representation from corresponding stocks' information as following two steps.

\subsubsection{Initializing the Predefined concepts' Representations}
\label{subsubsec:initialpredefined}
\begin{figure}[t]
	\begin{center}
		\includegraphics[width=1\columnwidth]{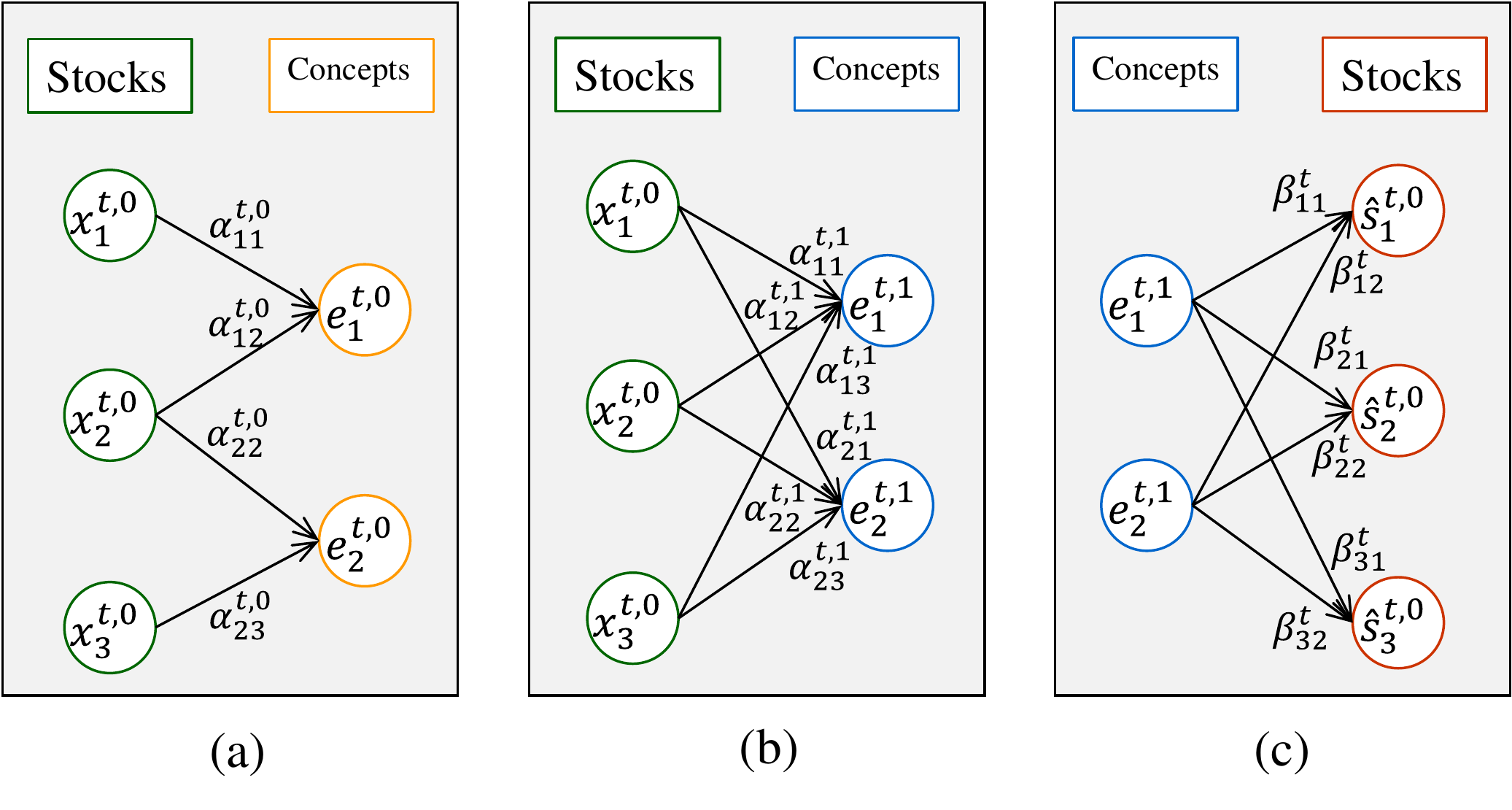}
	\end{center}
	\vspace{-0.2in}
	\caption{(a) Initializing the predefined concepts' representations (Section~\ref{subsubsec:initialpredefined}); (b) Correcting the predefined concepts' representations (Section~\ref{subsubsec:exctrapredefined}); (c) Aggregating the concepts' shared information to the stocks (Section~\ref{subsec:Aggregating})}
	\vspace{-0.15in}
	\label{fig:predefined_concept}
\end{figure}

We initialize a predefined concept's representation with the weighted element-sum of stock embeddings under this concept.
Since not all stocks contribute equally to a concept, referring to the calculation of the stock market index, we use the market capitalization of stock as the contribution weight from a stock to a predefined concept. 
More specifically, the weight from the stock $i$ to the predefined concept $T_k$ is: 
\begin{equation}
	\alpha^{t,0}_{ki} = \dfrac{c^t_{i}}{\sum_{j \in \mathcal{N}_k^t} c^t_{j}},
\end{equation}
where $c^t_{i}$ is the market capitalization of stock $i$ at date $t$, and $\mathcal{N}_k^t$ is the set of stocks that related to the concept $T_k$.
Then we aggregate the embeddings of stocks under the same concept $T_k$ to the concept $T_k$ with the weight $\alpha^{t,0}_{ki}$, and $e^{t,0}_k$ is the initial representation of the predefined concept $T_k$:
\begin{equation}
	e^{t,0}_k= \sum_{i \in \mathcal{N}_k^t} \alpha^{t,0}_{ki} x^{t,0}_i.
\end{equation}
Figure~\ref{fig:predefined_concept} (a) is an example of initializing representations of the predefined concepts when stock $1$ and $2$ share the predefined concept $T_1$ and stock $2$ and $3$ share the predefined concept $T_2$.

\subsubsection{Correcting the Predefined concepts' Representations}
\label{subsubsec:exctrapredefined}
We further extract the information from the related stocks to the predefined concepts to correct concepts' representations. The design of the correcting process aims to solve two limitations of predefined concepts: 1) some of the predefined stock concepts are missing (these missing concepts nay be strongly correlated to individual stocks); 2) some concepts have little impact on the related stocks (these stocks should not be taken into considerate when process concepts' representations). 
We correct the predefined concepts' representations based on the similarities between all stocks and all concepts to address the two limitations.
Intuitively, 1) if a stock is highly similar to a concept while the stock is not related to the concept, we assume that this concept is a missing concept for the stock.
On the contrary, 2) if a stock owns a low similarity with a concept while the stock is connected to the concept, we assume that this concept is not essential.

To describe the degree of connection between stocks and concepts, we firstly compute the cosine similarity $v_{ki}^{t,0}$ between each stock embedding $x_i^{t,0}$ and each predefined concept $T_k$'s initial representation $e^{t,0}_k$. Then we normalize the cosine similarity $v_{ki}^{t,0}$ using the softmax function and obtain the aggregated weight $\alpha^{t,1}_{ki}$ as following equation, 

\begin{equation}
	\begin{aligned}
		v^{t,0}_{ki} &= \mathrm{Cosine}(x_i^{t,0}, e^{t,0}_k)  =   \dfrac{x_i^{t,0} \cdot e^{t,0}_k}{|| x_i^{t,0} || \cdot || e^{t,0}_k||},\\
		\alpha^{t,1}_{ki} &=\frac{\exp \left(v^{t,0}_{ki}\right)}{\sum_{j \in \mathcal{S}^{t}} \exp \left(v^{t,0}_{kj}\right)}.
	\end{aligned}
\end{equation}
Finally, we use the aggregate weights to aggregate the stocks' embeddings to correct concepts' representations. The process is demonstrated in Equation~\ref{eq:correct_concept_pre}, where $W_e$ and $b_e$ are learnable parameters and LeakyReLU~\cite{maas2013rectifier} is the activation function. Notably, the use of aggregated weight $\alpha^{t,1}_{ki}$ in Equation~\ref{eq:correct_concept_pre} could solve the two limitations mentioned above.
\begin{equation}\label{eq:correct_concept_pre}
	\begin{aligned}
		e^{t,1}_k = \mathrm{LeakyReLU}&  \left(W_{e}\left(\sum_{i \in \mathcal{S}^{t}} \alpha^{t,1}_{ki} x^{t,0}_i\right) + b_{e}\right).
	\end{aligned}
\end{equation}
Figure~\ref{fig:predefined_concept} (b) is the illustration of correcting the shared information of predefined concepts.

\subsection{Extracting the Hidden Concepts' Shared Information}
\label{subsec:extrahidden}
\begin{figure}[t]
	\begin{center}
		\includegraphics[width=0.8\columnwidth]{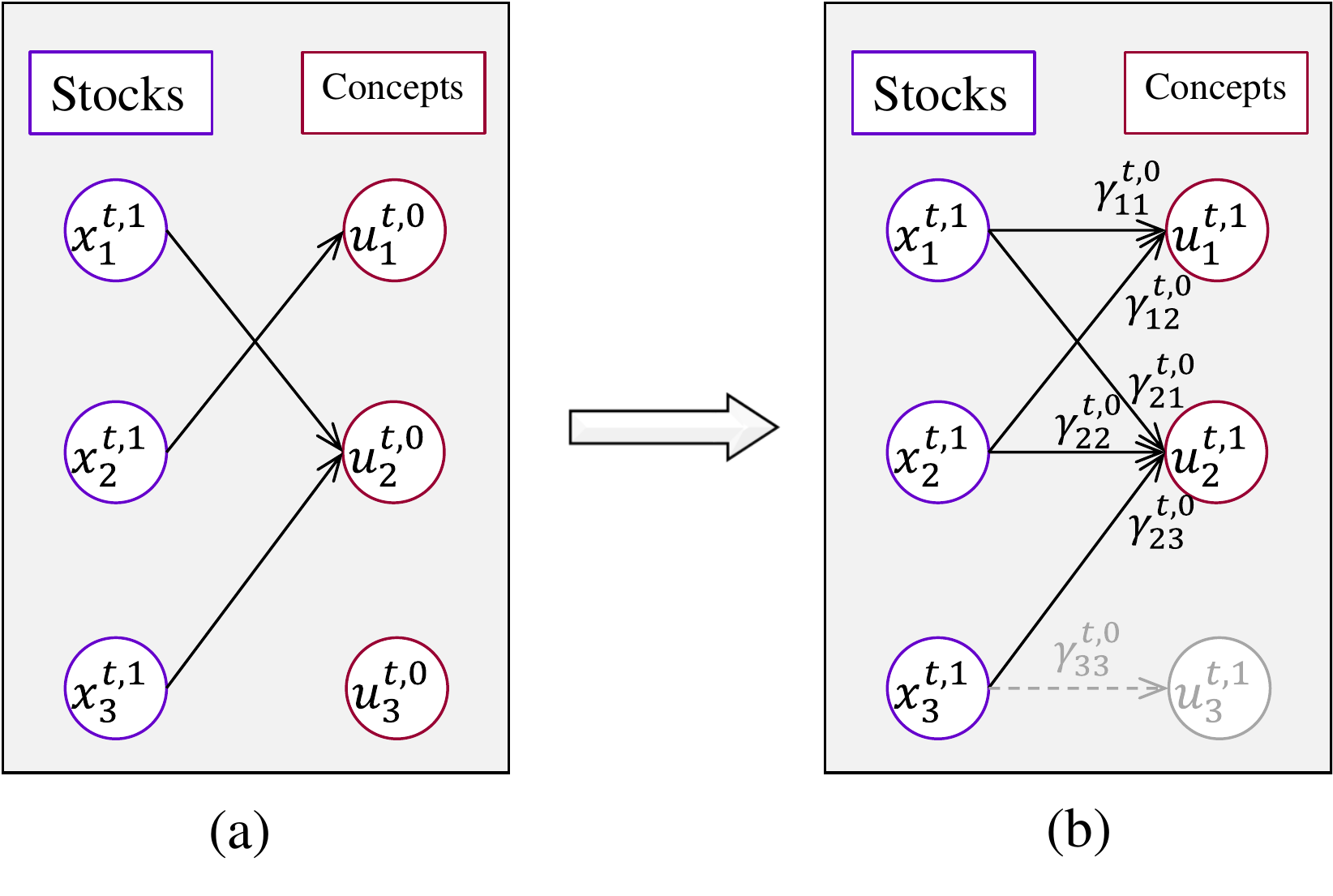}
	\end{center}
	\vspace{-0.2in}
	\caption{Illustration of extracting the hidden concepts' shared information (Section~\ref{subsec:extrahidden}).}
	\vspace{-0.2in}
	\label{fig:hidden_concept}
\end{figure}

In addition to the predefined stock concepts, some hidden stock concepts are not discovered and defined by human experts.
In the hidden concept module, we utilize a simple, effective algorithm to extract the hidden concepts and corresponding representations. Similar to the predefined concept, the extracted hidden concepts' representations contain the shared information of stocks related to the same hidden concepts. 
As described in Section~\ref{sec:overview}, the input of hidden concept module $x_t^{t, 1}$ remove the effect of shared information $\hat{x}^{t, 0}_i$ of predefined concepts from the stock embeddings $x_i^{t, 0}$.
The algorithm can be summarized as following steps:

\begin{enumerate}[leftmargin=1.5em]
	\item Initializing the hidden concepts' representations. We assume that there are $n$ hidden concepts, which is corresponding to the $n$ stocks. We use the stock $i$'s embedding $x_i^{t, 1}$ to initialize the corresponding hidden concept $H_i$'s embedding $u_i^{t, 0}$. 
	\item Computing the cosine similarity between all stocks and all hidden concepts with following equation:
	\begin{equation}
		\gamma_{ki}^{t,0} = \mathrm{Cosine}(x_i^{t,1}, u^{t,0}_k)  =   \dfrac{x_i^{t,1} \cdot u^{t,0}_k}{|| x_i^{t,1} || \cdot || u^{t,0}_k||},
	\end{equation}
	where $\gamma_{ki}^{t,0}$ is the cosine similarity between the stock $i$ and the hidden concept $H_k$.
	\item Connecting stocks with hidden concepts. We connect each stock with the most similar hidden concept except for its hidden concept (initialized by this stock's embedding) and delete the hidden concepts that do not connect with any stocks. As shown in Figure~\ref{fig:hidden_concept} (a), the most similar concept of the stock embeddings $x_1^{t,1}$, $x_2^{t,1}$ and $x_3^{t,1}$ is the hidden concept $H_2$, $H_1$ and $H_2$, respectively, and the hidden concept $H_3$ will be deleted because it does not connect with any stocks.
	\item Adding connections between stocks with its own hidden concept if its concept is not deleted. Take Figure~\ref{fig:hidden_concept} (b) as an example, the stock embeddings $x_1^{t,1}$ and $x_2^{t,1}$ are connected to concepts $H_1$ and $H_2$, respectively.
	\item Obtaining the representations of hidden concepts. We utilize the cosine similarity $\gamma^{t,0}$ to aggregated information from stocks to hidden concepts, and then we get the hidden concepts' representations $u^{t,1}_k$ of hidden concept $H_k$:
	\begin{equation}
		u^{t,1}_k=\mathrm{LeakyReLU} \left(W_{u}\left( \sum_{i\in\mathcal{M}_k^t} \gamma^{t,0}_{ki} x^{t,1}_i\right) + b_{u}\right),
	\end{equation}
	where $\mathcal{M}_k^t$ is the set of stocks that connect to the concept $H_k$, $W_u$ and $b_u$ are learnable parameters and LeakyReLU is the activation function. 
	
\end{enumerate}

\subsection{Aggregating the Concepts' Shared Information to Stocks}
\label{subsec:Aggregating}
In both the predefined concept module and hidden concept module, we aggregate the concepts' representations to the stocks for acquiring the shared information of stocks with the same concepts.
Since the shared information on different concepts has different importance for the stocks, some concepts may have higher importance while others may have lower; we apply the attention mechanism to learn the importance of each concept for a stock.

In the predefined concept module, we first compute the cosine similarity between the stock embedding $x_i^{t, 0}$ and the predefined concept $T_k$'s representation $e^{t,1}_k$, and then normalize the similarity value $v^{t, 1}_{ik}$ across all concepts in the predefined concept set $\mathcal{G}^t$ using the softmax function:
\begin{equation}
	\begin{aligned}
		v^{t, 1}_{ik} &= \mathrm{Cosine}(x^{t,0}_i, e^{t,1}_k)  =   \dfrac{x^{t,0}_i \cdot e^{t,1}_k}{|| x^{t,0}_i || \cdot || e^{t,1}_k ||},\\
		\beta^{t}_{ik} &= \dfrac{\exp \left(v^{t, 1}_{ik}\right)}{\sum_{j \in \mathcal{G}^t} \exp \left(v^{t, 1}_{ij}\right)},
	\end{aligned}
\end{equation}
thus we can obtain the aggregated weight $\beta^t_{ik}$ from concept representations $e_k^{t,1}$ to the stock $i$. We feed the aggregated information $\sum_{k \in \mathcal{G}^t} \beta^t_{ik} e^{t,1}_k$ into a fully-connected layer with LeakyReLU activation function, and obtain the shared information of the stocks correlated with the predefined concepts at date $t$:
\begin{equation}
	\hat{s}_i^{t, 0} =  \mathrm{LeakyReLU} \left(W_{s}^0\left(\sum_{k \in \mathcal{G}^t} \beta^t_{ik} e^{t,1}_k\right) + b_{s}^0\right).
	\label{equ:es}
\end{equation}
Figure~\ref{fig:predefined_concept} (c) illustrates an example of aggregating the concepts' shared information to the stocks. For the hidden concept module, the process of aggregating the hidden concepts' shared information is the same as the predefined concept module.

\subsection{Output of Predefined/Hidden Concept Module}
\label{subsec:output}
In predefined concept module, we feed the $\hat{s}_i^{t, 0}$ into two fully-connected layers with LeakyReLU activation function to generate the backcast and forecast outputs of predefined concept module:
\begin{equation}
	\begin{aligned}
		\hat{x}^{t, 0}_i&=\mathrm{LeakyReLU}\left(W_b^0 \hat{s}_i^{t, 0} + b_b^0\right),\\ 
		\hat{y}^{t, 0}_i &=\mathrm{LeakyReLU}\left(W_f^0 \hat{s}_i^{t, 0}+ b_f^0\right) ,
	\end{aligned}
\end{equation}
where $\hat{x}^{t, 0}_i$ and $\hat{y}^{t, 0}_i$ are the backcast and forecast outputs of predefined concept module.
Same as the design in the predefined concept module, the hidden concept module also has two output branches: backcast output  $\hat{x}^{t, 1}_i$ and the forecast output $\hat{y}^{t, 1}_i$.

\subsection{Individual Information Module}
\label{subsec:individual}
Beside the shared information of stocks connected with the same predefined and hidden concepts, each stock's individual information is also essential for stock trend forecasting. 
Therefore, we utilize the individual information module for mining the residual individual information of stocks.
The input $x_i^{t, 2}$ of individual information module further subtract the effect of hidden concepts’ shared information $\hat{x}^{t, 1}_i$.
We feed the input $x_i^{t, 2}$ into a fully connected layer with the LeakyReLU activation function to generate the forecast output of individual information module:
\begin{equation}
	\hat{y}^{t, 2}_i =\mathrm{LeakyReLU}\left(W_f^2 x_i^{t, 2}+ b_f^2\right) ,
\end{equation}
The forecast output $\hat{y}^{t, 2}_i $ represents the individual information of stock $i$ that removes the effect of stock shared information.

\subsection{Training Objective Function}
\label{sec:forecasting}
We leverage the Adam algorithm~\cite{kingma:adam} to optimize our HIST framework by minimizing the mean squared error (MSE) loss function:
\begin{equation}
	\begin{split}
		\mathcal{L} =  \sum_{t \in \mathcal{T}} \mathrm{MSE}\left(p^t, d^t\right)  = \sum_{t \in \mathcal{T}} \sum_{i \in \mathcal{S}^{t}} \frac{\left( p_i^t - d_i^t \right)^2}{|\mathcal{S}^{t} |},
	\end{split}
\end{equation}
where $\mathcal{T}$ is the set of dates in the training set and the $\mathcal{S}^{t}$ is the set of stocks in date $t$; the $p_i^t$ and $d_i^t$ are the stock trend prediction and the ground truth stock trend of stock $i$ at date $t$, respectively. 

%% file: experiment.tex
\section{Experiments}
\label{sec:experiment}
In this section, we conduct experiments for our HIST framework, aiming to answer the following research questions:
\begin{itemize}[leftmargin=1.5em]
	\item \textbf{RQ1}: How does our model perform compared with existing stock trend forecasting methods?
	\item \textbf{RQ2}: What is different components’ effect in our framework?
	\item \textbf{RQ3}: Can our model achieve a higher investment return in the investment simulation on real-world datasets?
	\item \textbf{RQ4}: What hidden concepts' shared information we mined?
\end{itemize}

\begin{table*}[t]
	\label{tab:main_results}
	\centering
	\caption{The main results (and its standard deviation) on CSI 100 and CSI 300.}
	\vspace{-0.1in}
	\resizebox{0.90\textwidth}{!}{
		\begin{tabular}{l|cc|cccc|cc|cccc}
			\hline
			\\[-1em]
			\multirow{3}{*}{Methods} & \multicolumn{6}{c|}{CSI 100} & \multicolumn{6}{c}{CSI 300}	\\
			\\[-1em]
			\cline{2-13} 
			\\[-1em]
			& \multirow{2}{*}{IC} &\multirow{2}{*}{Rank IC} & \multicolumn{4}{c|}{Precision@N ($\uparrow$)}& \multirow{2}{*}{IC} &\multirow{2}{*}{Rank IC} & \multicolumn{4}{c}{Precision@N ($\uparrow$)} \\
			\\[-1em]
			&  ($\uparrow$)& ($\uparrow$) & 3 & 5 & 10& 30& ($\uparrow$) & ($\uparrow$)& 3 & 5 & 10& 30\\
			\\[-1em]
			\hline
			\\[-1em]
			\multirow{2}{*}{MLP}&0.071&0.067&56.53& 56.17  & 55.49 & 53.55 & 0.082& 0.079& 57.21  & 57.10  & 56.75& 55.56 \\ 
			&(4.8e-3)&(5.2e-3)&(0.91)&(0.48)&(0.30)&(0.36)&(6e-4)&(3e-4)&(0.39)&(0.33)&(0.34)&(0.14)\\
			\\[-1em]
			\hline
			\\[-1em]
			\multirow{2}{*}{LSTM~\cite{hochreiter1997long}}& 0.097& 0.091& 60.12  & 59.49  & 59.04 & 54.77 & 0.104&  0.098& 59.51  & 59.27  & 58.40 & 56.98 \\ 
			&(2.2e-3)&(2.0e-3)&(0.52)&(0.19)&(0.15)&(0.11)&(1.5e-3)&(1.6e-3)&(0.46)&(0.34)&(0.30)&(0.11)\\
			\\[-1em]
			\hline
			\\[-1em]
			\multirow{2}{*}{GRU~\cite{chung2014empirical}} & 0.103& 0.097& 59.97  & 58.99  & 58.37 & 55.09 & 0.113& 0.108& 59.95  & 59.28  & 58.59 & 57.43 \\ 
			&(1.7e-3)&(1.6e-3)&(0.63)&(0.42)&(0.29)&(0.15)&(1e-3)&(8e-4)&(0.62)&(0.35)&(0.40)&(0.28)\\
			\\[-1em]
			\hline
			\\[-1em]
			\multirow{2}{*}{SFM~\cite{zhang2017stock}}  &0.081&0.074&57.79&56.96&55.92&53.88&0.102&0.096&59.84&58.28&57.89&56.82\\
			&(7.0e-3)&(8.0e-3)&(0.76)&(1.04)&(0.60)&(0.47)&(3.0e-3)&(2.7e-3)&(0.91)&(0.42)&(0.45)&(0.39)\\
			\\[-1em]
			\hline
			\\[-1em]
			\multirow{2}{*}{GATs~\cite{velickovic2018graph}}& 0.096&0.090& 59.17 & 58.71  & 57.48 &54.59  & 0.111& 0.105& 60.49  & 59.96  & 59.02 & 57.41 \\ 
			&(4.5e-3)&(4.4e-3)&(0.68)&(0.52)&(0.30)&(0.34)&(1.9e-3)&(1.9e-3)&(0.39)&(0.23)&(0.14)&(0.30)\\
			\\[-1em]
			\hline
			\\[-1em]
			\multirow{2}{*}{ALSTM~\cite{feng2019enhancing}}&0.102 &0.097&60.79  & 59.76  &58.13  & 55.00 &0.115& 0.109&59.51  & 59.33   & 58.92 & 57.47 \\ 
			&(1.8e-3)&(1.9e-3)&(0.23)&(0.42)&(0.13)&(0.12)&(1.4e-3)&(1.4e-3)&(0.20)&(0.51)&(0.29)&(0.16)\\
			\\[-1em]
			\hline
			\\[-1em]
			\multirow{2}{*}{Transformer~\cite{ding2020hierarchical}}&0.089  &0.090& 59.62&59.20 &57.94&54.80&0.106&0.104&60.76&60.06&59.48&57.71\\ 
			&(4.7e-3)&(5.1e-3)&(1.20)&(0.84)&(0.61)&(0.33)&(3.3e-3)&(2.5e-3)&(0.35)&(0.20)&(0.16)&(0.12)\\
			\\[-1em]
			\hline
			\\[-1em]
			\multirow{2}{*}{ALSTM+TRA~\cite{lin2021learning}}& 0.107 &0.102&60.27&59.09&57.66& 55.16&0.119&0.112&60.45 & 59.52 &59.16 &58.24 \\ 
			&(2.0e-3)&(1.8e-3)&(0.43)&(0.42)&(0.33)&(0.22)&(1.9e-3)&(1.7e-3)&(0.53)&(0.58)&(0.43)&(0.32)\\
			\\[-1em]
			\hline
			\\[-1em]
			\multirow{2}{*}{\textbf{HIST}}	&\textbf{0.120} &\textbf{0.115}& \textbf{61.87} & \textbf{60.82} & \textbf{59.38}& \textbf{56.04}&\textbf{0.131}& \textbf{0.126}& \textbf{61.60} & \textbf{61.08} & \textbf{60.51}& \textbf{58.79}\\ 
			&\textbf{(1.7e-3)}&\textbf{(1.6e-3)} &\textbf{(0.47)}  & \textbf{(0.43)} & \textbf{(0.24)}&\textbf{(0.19)} &\textbf{(2.2e-3)}& \textbf{(2.2e-3)}&\textbf{(0.59)}& \textbf{(0.56)} & \textbf{(0.40)}& \textbf{(0.31)}
			\\[-1em]
			\\ 
			\hline
		\end{tabular}
	}
	\vspace{-0.1in}
\end{table*}

\subsection{Datasets}
\xhdr{Stock Sets} 
We evaluate our HIST framework on the stocks of two popular and representative stock sets: CSI 100 and CSI 300. CSI 100 and CSI 300 comprise the largest $100$ and $300$ stocks in the China A-share market, respectively. The CSI 100 reflects the performances of most influential large-cap A-shares market, and the CSI 300 reflects the overall performance of China A-share market.

\xhdr{Stock Features}
We use the stock features of Alpha360 in the open-source quantitative investment platform Qlib\footnote{\url{https://github.com/microsoft/qlib}}~\cite{yang2020qlib}.
The Alpha360 dataset contains $6$ stock data on each day, which are \textit{opening price}, \textit{closing price}, \textit{highest price}, \textit{lowest price}, \textit{volume weighted average price} (VWAP) and \textit{trading volume}. 
For each stock on date $t$, Alpha360 looks back $60$ days to construct a $360$-dimensional historical stock data as a stock feature of this stock at date $t$. 
We use the features of stocks in CSI 100 and CSI 300 both from 01/01/2007 to 12/31/2020, and split them by time to obtain training set (from 01/01/2007 to 12/31/2014), validation set (from 01/01/2015 to 12/31/2016), and test set (from 01/01/2017 to 12/31/2020). 
We use the stock trend defined in Equation~\ref{eq:stock-trend} as the label for each stock on each date and apply normalization on labels of the same date.

\xhdr{Predefined Concept Data}
We collect two types of essential predefined concept data: the industry and the main business of stocks\footnote{We collect the industry and main business data from Tushare: \url{https://tushare.pro/}.}. 
Due to the change of stocks in the stock set (the stocks in CSI 100 and CSI 300 will change half a year according to their current market capitalization) and the change of stocks' industry and businesses, the number of predefined concepts may change dynamically.
The average number of predefined concepts in each day is 410 in CSI 100 and 1344 in CSI 300.

\subsection{Experimental Setting}
\xhdr{Baselines}
We compare our HIST framework with the following methods: MLP, LSTM~\cite{hochreiter1997long}, GRU~\cite{chung2014empirical}, SFM~\cite{zhang2017stock}, GATs~\cite{velickovic2018graph}, ALSTM~\cite{feng2019enhancing}, Transformer~\cite{ding2020hierarchical} and ALSTM+TRA~\cite{lin2021learning}. The detailed introduction of these baselines are in Appendix~\ref{appendix:baseline}.

\xhdr{Evaluation Metrics}
We first use two widely-used evaluation metrics: the \textbf{Information Coefficient (IC)}~\cite{lin2021learning} and \textbf{Rank IC}~\cite{li2019individualized}.
Besides, we also use the \textbf{Precision@N}, where N is 3, 5, 10, and 30, to evaluate the precision of the top N predictions of each model.
The detailed introduction of each evaluation metric is described in Appendix~\ref{appendix:metrics}.
To eliminate the fluctuations caused by different initialization, we repeat the training and testing procedure 10 times for all results and report the average value and standard deviation.

\xhdr{Implementation Details}
We implement our framework with the PyTorch library\footnote{The source code of our method and all the other baselines are available at this repository: \url{https://github.com/Wentao-Xu/HIST}.}~\cite{paszke2019pytorch}, and run all experiments on a single NVIDIA Tesla V100 GPU. The hyper-parameters setting of our method and the baselines are in Appendix~\ref{appendix:hyper-para}.

\subsection{Main Results (RQ1)}
Table $1$ shows the experimental results of HIST and other baselines on the stocks of CSI 100 and CSI 300. Our HIST framework achieves the highest IC, Rank IC, and Precision@N. 
Our proposed HIST framework can achieve better results than some latest stock trend forecasting methods like ALSTM, Transformer, and ALSTM+TRA; thus, our framework is more effective than existing stock trend forecasting methods.
Besides, although the GATs can also capture the cross-stock connections, it only utilizes stationary cross-stock relations. 
Compared with GATs, our HIST can model the dynamic relevance degree between stocks and concepts; thus, it can capture the temporal and complicated cross-stock relations between stocks. 
Moreover, our method can further mine the shared information of hidden concepts, so our HIST outperforms existing GNN-based cross-stock technical analysis method GATs.

\begin{table*}[t]
	\centering
	\caption{The results of ablation study. In this table, the \textit{Initialize} and \textit{Correct} are the initialization and correction of predefined concepts in predefined concept module (corresponding to Section~\ref{subsubsec:initialpredefined} and~\ref{subsubsec:exctrapredefined}), respectively. The Hidden represents the hidden concept module, and the Individual indicates the individual information module. The \cmark and - indicate having or not having the component in the variants. The Precision is the average of different Precision@N values where N are 3,5,10 and 30.}
	\vspace{-0.1in}
		\resizebox{0.8\textwidth}{!}{
	\begin{tabular}{ c c c c |  c  c c |  c c c}
		\hline
		\\[-1em]
		\multicolumn{2}{c}{Predefined} &\multirow{2}{*}{Hidden} &\multirow{2}{*}{Individual}  &  \multicolumn{3}{c|}{CSI 100} &   \multicolumn{3}{c}{CSI 300} \\
		\textit{Initialize}&\textit{Correct}&&  	&  IC $(\uparrow)$ & Rank IC $(\uparrow)$	&Precision $(\uparrow)$ &  IC $(\uparrow)$& Rank IC $(\uparrow)$   &Precision $(\uparrow)$\\ 
		\\[-1em]
		\hline
		\\[-1em]
		\cmark&-&-& -& 0.087 &  0.084 &57.40 & 0.101 &  0.094&57.89\\
		\cmark&\cmark&-& -& 0.099 &  0.096&58.14 & 0.112 &  0.106& 58.74\\
		-&-&\cmark& -& 0.097 &  0.096  & 58.05& 0.110 &  0.104&58.46\\
		\cmark&\cmark&\cmark& - & 0.111 &0.107& 58.76& 0.120 &  0.113&59.55\\
		\cmark&\cmark&\cmark& \cmark& \textbf{0.120} & \textbf{0.115} &\textbf{59.53}& \textbf{0.131} &  \textbf{0.126}&\textbf{60.50}\\
		\\[-1em]
		\hline
	\end{tabular}}
	\label{tab:ablation}
	\vspace{-0.15in}	
\end{table*}
\subsection{Ablation Study (RQ2)}
We apply an ablation study on our framework to study the effect of different modules in our framework.
Specifically, we study the effect of initializing and correcting the predefined concept in the predefined concept module and the impact of the hidden concept module and individual information module.
We study these components' effects by removing some components and observing the new experimental results.
For example, when we only use the predefined concept module, we directly utilize the output of the predefined concept module to forecast the stock price trend.
Table~\ref{tab:ablation} shows the results of ablation study, and we have the following observations:
\begin{enumerate}[leftmargin=1.5em]
	\item[{(1)}]  We can find that correcting the predefined concept can improve the performance, which implies that correcting the shared information of predefined concepts, mining the missing stock concepts, and reducing the effect of concepts with less significance can improve the performance of our framework.
	\item[{(2)}]  Removing the predefined concept module or hidden concept module would reduce the performance, so the shared information of predefined and hidden concepts are both vital, and we can not ignore any one of them. 
	\item[{(3)}] Removing the individual information module would also reduce the performance, so the individual information of each stock is also crucial for stock trend forecasting.
\end{enumerate}

\begin{figure}[t]
	\centering
	\includegraphics[width=0.96\columnwidth]{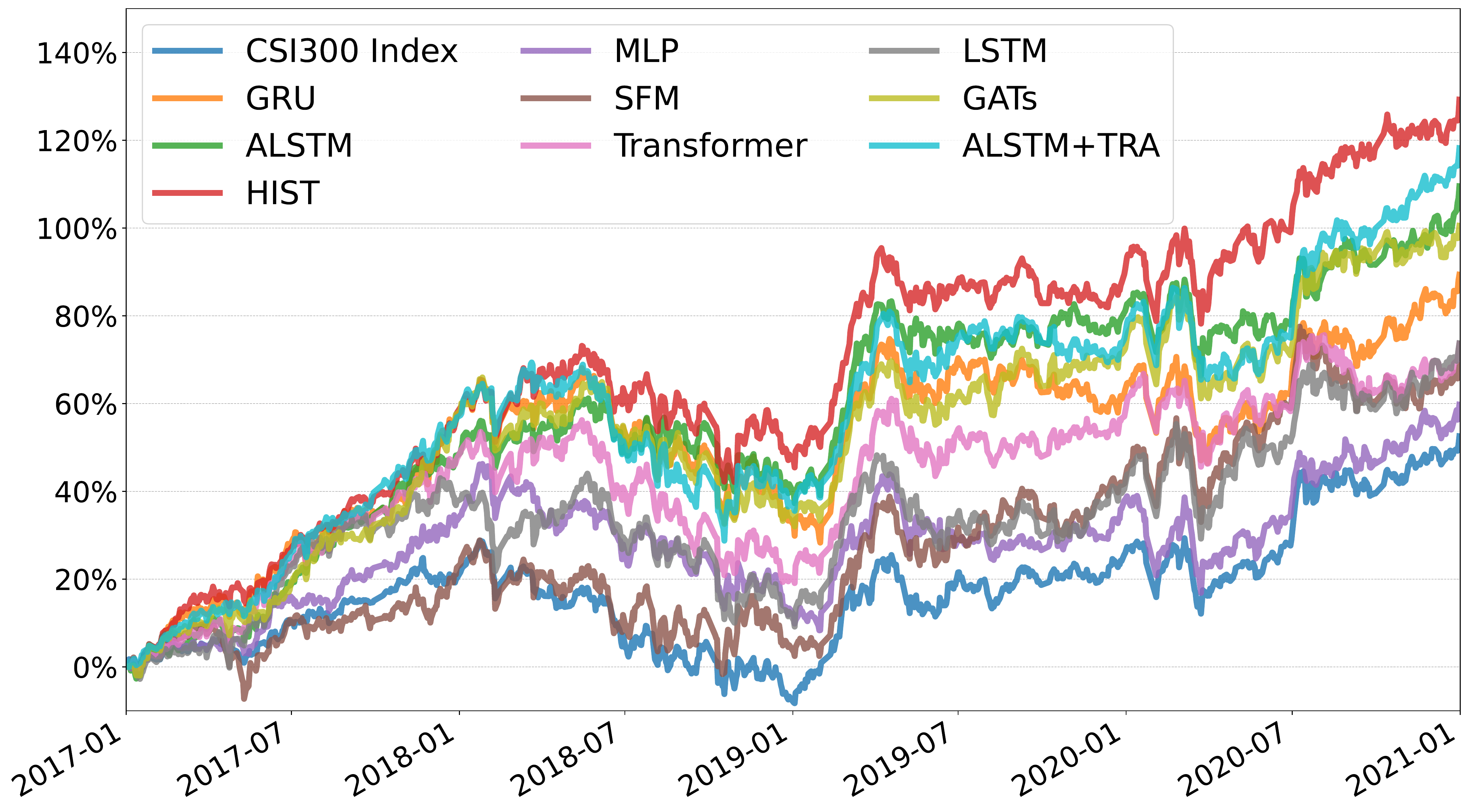}
	\vspace{-.1in}
	\caption{Cumulative Return on CSI 300 from 2017 to 2020.}
	\vspace{-.2in}	
	\label{fig:return}
\end{figure} 

\subsection{Investment Simulation (RQ3)}
To further evaluate our HIST framework's effectiveness, we utilize an investment strategy to simulate the investment in the CSI 300's test set (from 01/01/2017 to 12/31/2020). To be specific, we rank the stocks on date $t$ from high to low according to the stock trend's predictions, then select the top $k$ stocks to evenly invest and sell the currently held stocks that not in the top $k$. To simulate real-world trading, we assume the initial account capital is $1 e^8$, and we consider a transaction cost of $0.05\%$ for buying shares and $0.15\%$ for selling shares. We use the Cumulative Return to evaluate the investment simulation result:
\begin{itemize}[leftmargin=1.5em]
	\item \textbf{Cumulative Return (CR)} is a the aggregate amount that the investment has gained or lost over time, and the cumulative return is calculated by: ${\rm CR} = \dfrac{\rm current\ capital - initial\ captial}{\rm initial\ captial}$.
\end{itemize}
To find the best selection of number $k$, we conduct a grid search to find the best value that maximizes the Cumulative Return on the validation set. We tune the $k\in \{10, 20, 30, 40, 50\}$,  and we find that we can achieve the highest return when $k$ is 30.
Figure~\ref{fig:return} shows the results of the Cumulative Return. Despite the stock market crash in 2018, our HIST framework still can gain a over 120\% Cumulative Return from 2017 to 2020, and 70\% better than the CSI 300 index.
Figure~\ref{fig:return} illustrates that our HIST framework is not only outperforming some state-of-the-art methods, such as Transformer and ALSTM+TRA but also better than some graph-based models like GATs.

\subsection{Analyzing the Hidden Concepts (RQ4)}
\begin{figure}[t]
	\centering
	\includegraphics[width=0.9\columnwidth]{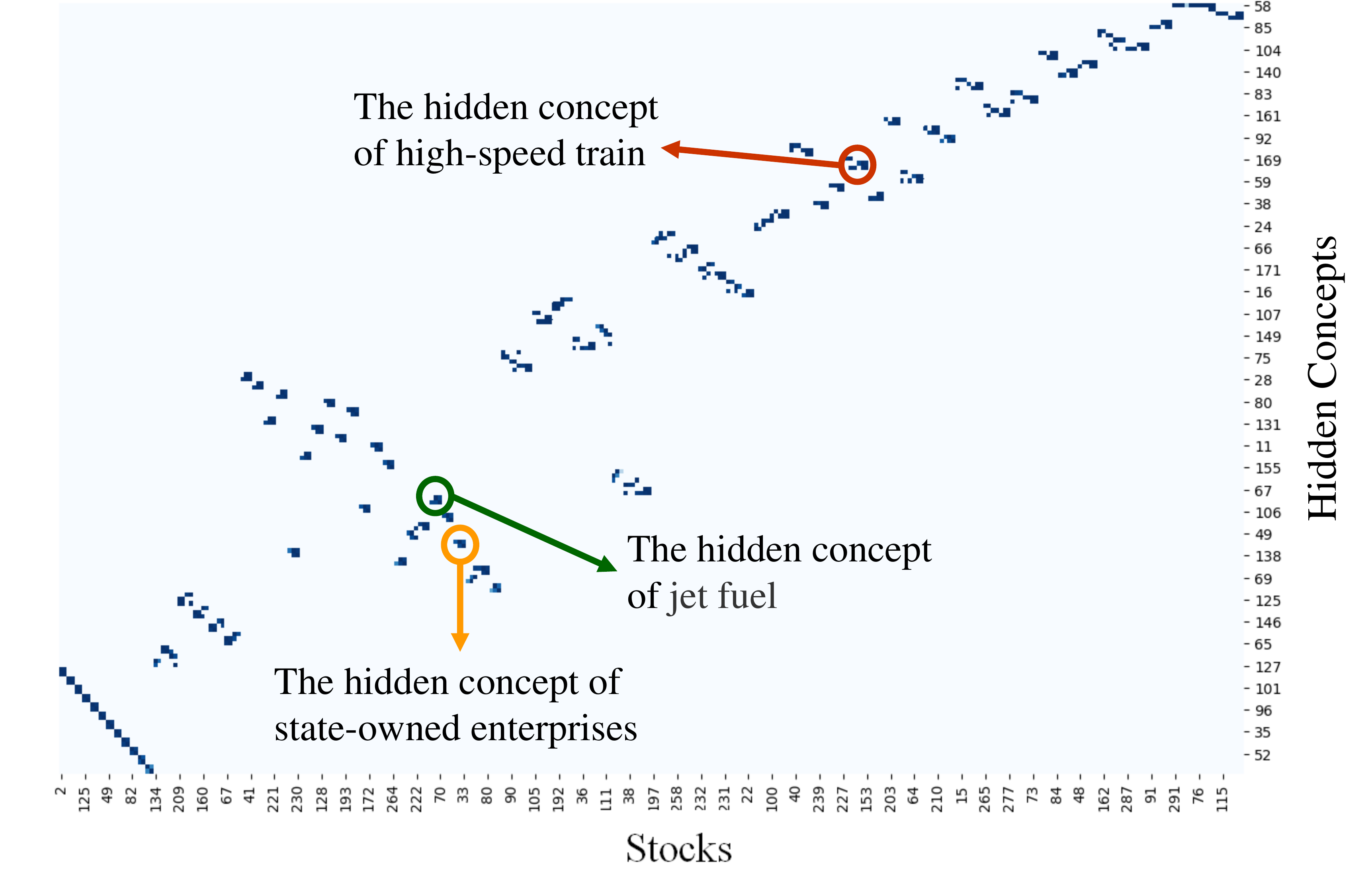}
	\vspace{-0.1in}
	\caption{Visualization of the mapping between stocks and hidden concepts (a hierarchically-clustered heatmap on the weights matrix $\gamma^{t,0}$ in Section~\ref{subsec:extrahidden}) on CSI 300 at Jan. 3, 2017. The row index indicates the stock and the column index indicates the hidden concepts, and a point $(i, j)$ in this figure represents the stock $i$ connects with the hidden concept $H_j$.}
	\vspace{-0.15in}
	\label{fig:visualization}
\end{figure}

We visualize the weights matrix $\gamma^{t,0}$ in the hidden concept module to analyze the hidden concepts we mined.
Figure~\ref{fig:visualization} is the hierarchically clustered heatmap on the matrix $\gamma^{t,0}$, displaying the relationships between stocks and hidden concepts. 
Observing the specific stocks connected with the same hidden concept can know which stocks share the same hidden information.
When we enlarge Figure~\ref{fig:visualization}, we find the 4 stocks in the red circle include the railway construction company, the rolling stock manufacture company, and the high-speed train maintenance company, so they have the same hidden concept: `high-speed train.'
Similarly, the 3 stocks in the green circle can also have the same hidden concept `jet fuel' because they are petrochemical companies or airlines.
The predefined concepts we used do not cover these hidden concepts we mined.

%% file: conclusion.tex
\section{Conclusion and Future Work}
\label{sec:conclusion}
We propose a graph-based framework, HIST, that can effectively mine the concept-oriented shared information among stocks and significantly improve stock trend forecasting performance. Specifically, after extracting the temporal features of stocks using GRU, we sequentially apply three modules that can extract the shared information based on the predefined concepts and the mined hidden concepts, as well as individual features that are not captured by shared features. Experimental results in the real-world stock market demonstrated the effectiveness of our proposed framework.

In the future, we plan to mine more abundant and diverse stock shared information from the Web, such as from news, social media and discussion board, and fuse different kinds of stock shared information to forecasting the stock price trend.  

%% file: appendix.tex
\appendix

\section{Details of Experimental Setting}

\subsection{Baselines} 
\label{appendix:baseline}
We compare our proposed HIST framework with the following stock trend forecasting methods:
\begin{itemize}[leftmargin=1.5em]
	\item \textbf{MLP}: a $3$-layers multi-layer perceptron (MLP) with the number of units on each layer is $512$.
	\item \textbf{LSTM}~\cite{hochreiter1997long}: a Long Short-Term Memory (LSTM) network based stock trend forecasting method.
	\item \textbf{GRU}~\cite{chung2014empirical}: a Gated Recurrent Unit (GRU) network based  stock trend forecasting method.
	\item \textbf{SFM}~\cite{zhang2017stock}: a RNN that decomposes the hidden states into multiple frequency components to model multi-frequency patterns.
	\item \textbf{GATs}~\cite{velickovic2018graph}: a forecasting model that utilizes graph attention networks (GATs) to aggregate stock embeddings encoded by GRU on the stock graph. We use the stocks as nodes to construct a stock graph, and two stocks have a relation when they share the same predefined concept.
	\item \textbf{ALSTM}~\cite{feng2019enhancing}: a variant of LSTM with a temporal attentive aggregation layer to aggregate information from all hidden states in previous timestamps.
	\item \textbf{Transformer}~\cite{ding2020hierarchical}: A transformer~\cite{vaswani2017attention} based stock trend forecasting model.
	\item \textbf{ALSTM+TRA}~\cite{lin2021learning}: an ALSTM extension that uses Temporal Routing Adaptor (TRA) to model multiple trading patterns.
\end{itemize}

\subsection{Evaluation Metrics}
\label{appendix:metrics}
We first use two widely used indicators in the quantitative investment domain as evaluation metrics: the \textbf{Information Coefficient (IC)}~\cite{lin2021learning} and \textbf{Rank IC}~\cite{li2019individualized}.
The IC is the Pearson correlation coefficient~\cite{benesty2009pearson} between the labels and predictions. 
The Rank IC is the Spearman's rank correlation coefficient, calculated by: $\mathrm{Rank\ IC}(y^t, \hat{y}^t) = \mathrm{corr}(rank_{y}^t, rank_{\hat{y}}^t)$, where $\mathrm{corr}(\cdot)$ is the Pearson correlation coefficient. 
We sort the stock trend labels and predictions of all stocks on each day from high to low and acquire the ranks $rank_{y}^t$ and $rank_{\hat{y}}^t$ of each stock's labels and predictions on the date $t$. 
We use the average IC and Rank IC of each day to evaluate the results of stock trend forecasting. 

Furthermore, for a stock trend forecasting model, the precision of its top N predictions is more vital for real-world stock investment. Therefore, we introduce another evaluation metric: \textbf{Precision@N}.
The Precision@N is the proportion of top N predictions on each day with the positive label. For example, when N is 10, and the labels of 5 among these top 10 predictions are positive, then the Precision@10 is 50\%.
We report the results of average Precision@N on each day, and we set the N as 3, 5, 10, and 30. 

\begin{table}[h]
	\caption{The selection of the hyper-parameters.}
	\vspace{-0.1in}
	\label{tab:hyper-para}
	\begin{tabular}{l|cc|cc}
		\hline
		\\[-1em]
		Hyper-parameters& \multicolumn{2}{c|}{Number of Units}  &  \multicolumn{2}{c}{Number of Layers} \\
		
		Dataset& CSI 100&  CSI 300 & CSI 100&  CSI 300  \\
		\\[-1em]
		\hline
		\\[-1em]
		MLP & 512 &512 & 3& 3 \\
		LSTM& 128&64&2 &  2\\
		GRU&128 & 64& 2& 2 \\
		SFM& 64& 128& 2& 2 \\
		GATs&128& 64 &2 &2  \\
		ALSTM&64 & 128 &2 & 2 \\
		Transformer&32 & 32&3 & 3 \\
		ALSTM+TRA&64 &128 & 2&2  \\
		\\[-1em]
		\hline
		\\[-1em]
		\textbf{HIST} &	128 &128 &2& 2 \\
		[-1em]\\
		\hline
	\end{tabular}
	\vspace{-0.1in}
\end{table}
\subsection{Hyper-parameters Settting}
\label{appendix:hyper-para}
The dimension $l$ of stock features in Alpha360 is $360$. and we set each training and testing batch as the stock features on the same date, so the batch size of our HIST framework and other baselines is equal to the number of stock on each day.

Besides, we tune our framework and other baselines using the grid search to select the optimal hyper-parameters based on the performance of validation set.
We search the number of hidden units $d$ of GRU in our stock features encoder, and the number of hidden units in other baselines in $\{32, 64, 128, 256, 512\}$; the number of layers of our stock feature encoder's GRU and other baselines in $\{1, 2, 3, 4\}$; the learning rate in $\{0.001, 0.0005, 0.0002, 0.0001\}$.
Table~\ref{tab:hyper-para} shows the selection of the number of units and layers in our HIST framework and other baselines, and the best learning rate in our framework and other baseline is $0.0002$.